\documentclass[amsmath,amssymb,twocolumn,superscriptaddress]{revtex4}
\usepackage{graphicx}% Include figure files
\usepackage{dcolumn}% Align table columns on decimal point

\def\be{\begin{equation}}
\def\ee{\end{equation}}
\def\beq{\begin{eqnarray}}
\def\eeq{\end{eqnarray}}

\usepackage{bm}% bold math
\usepackage{graphicx}
\usepackage{dcolumn}
\usepackage{amsmath}
\usepackage[latin1]{inputenc}
\usepackage{graphicx, psfrag}
\usepackage{amssymb}
\usepackage[colorlinks=true, citecolor=blue, urlcolor = blue, linkcolor= red, bookmarks=true]{hyperref}
\usepackage{float}
\usepackage{amsmath}
\usepackage{amsfonts}
\usepackage{dcolumn}
\usepackage{hyperref}
\usepackage{subfigure}
\usepackage{pgfplots}
\usepackage{epstopdf}
\usepackage{booktabs}

\begin{document}

\title{A power law solution for FRLW Universe with observational constraints} 

\author{Lokesh Kumar Sharma}
\email[Email:]{lokesh.sharma@gla.ac.in}
\affiliation{Department of Physics, GLA University, Mathura 281406, Uttar Pradesh, India}

\author{Suresh Parekh}
\email[Email:]{thesureshparekh@gmail.com}
\affiliation{Department of  Physics, SP Pune University, Pune 411007, Maharastra, India}

\author{Sanjay Maurya}
\email[Email:]{mau123sanjay@gmail.com}
\affiliation{Department of  Physics, SP Pune University, Pune 411007, Maharastra, India}

\author{Kuldeep Singh}
\email[Email:]{kuldeepbsingh573@gmail.com}
\affiliation{Department of  Physics, SP Pune University, Pune 411007, Maharastra, India}

\author{Saibal Ray}
\email[Email:]{saibal.ray@gla.ac.in }
\affiliation{Center for Cosmology, Astrophysics and Space Science (CCASS), GLA University, Mathura 281406, Uttar Pradesh, India} 

\author{Kalyani C.K. Mehta}
\email[Email:]{kalyani.c.k.mehta@gmail.com}
\affiliation{Department of Physics, Eberhard Karls University of Tubingen, Germany}

\author{Vaibhav Trivedi}
\email[Email:]{vbtrivedi06@gmail.com}
\affiliation{Department of Physics, Fergusson College, Pune 411004, Maharastra, India}

\date{\today}

\begin{abstract}
This paper examines a power law solution under $f(R,T)$ gravity for an isotropic and homogeneous universe by considering its functional form as $f(R,T) = R + \xi RT$, where $\xi$ is a positive constant. In $f(R,T)$ gravity, we have built the field equation for homogeneous and isotropic spacetime. The developed model's solution is $a = \alpha t^{\beta}$. We have used the redshift in the range $0 \leq z \leq 1.965$ and obtained the model parameters $\alpha$, $\beta$, $H_0$ by using the Markov Chain Monte Carlo (MCMC) method. The constrained values of the model parameter are as follows: $H_0 = 67.098^{+2.148}_{-1.792}$ km s$^{-1}$ Mpc$^{-1}$, $H_0 = 67.588^{+2.229}_{-2.170}$ km s$^{-1}$ Mpc$^{-1}$, $H_0 = 66.270^{+2.215}_{-2.181}$ km s$^{-1}$ Mpc$^{-1}$, $H_0 = 65.960^{+2.380}_{-1.834}$ km s$^{-1}$ Mpc$^{-1}$, $H_0 = 66.274^{+2.015}_{-1.864}$ km s$^{-1}$ Mpc$^{-1}$ which have been achieved by bounding the model with the Hubble parameter ($H(z)$) dataset, Baryon Acoustic Oscillations (BAO) dataset, Pantheon dataset, joint $H(z)$ + Pantheon dataset and collective $H(z)$ + BAO + Pantheon dataset, respectively. These computed $H_o$ observational values agree well with the outcomes from the Plank collaboration group. Through an analysis of the energy conditions' behaviour on our obtained solution, the model has been examined and analysed. Using the Om diagnostic as the state finder diagnostic tool and the jerk parameter, we have also investigated the model's validity. Our results show that, within a certain range of restrictions, the proposed model agrees with the observed signatures.

\end{abstract}

\maketitle

\section{Introduction}\label{1}
For an extended period, there has been a status quo on the spacetime of the Universe, which seems to be maintaining a non-dynamic phase until Hubble~\cite{Hubble1929} did not offer the concept of expanding Universe based on his observation of the galactic motion. However, after several decades, a new phase of the astrophysical observations, viz. high redshift supernovae~\cite{Riess1998,Riess2004}, $H(z)$ measurements of SN Ia~\cite{Perlmutter1998,Perlmutter1999,Bennett2003}, Cosmic Microwave Background Radiation (CMBR)~\cite{Spergel2003,Spergel2007}, baryon acoustic oscillations~\cite{Percival2010} and Planck data~\cite{Ade2014} have provided all strong shreds of evidence that we are now inhabiting not only in an expanding but also an accelerating phase of the universe~\cite{Aghanim2018}. 

The current Universe, however, was shown to have started in a decelerating phase. It has been a considerable issue for theoretical astrophysicists and cosmologists to understand the cause of the unexpected finding of the Universe's late-time acceleration~\cite{Mishra2019}. The anti-gravitational effects of non-baryonic energy are predicted by general relativity (GR) to be a significant contributor to the accelerated expansion of the current universe~\cite{Peebles2003}. This means that the cosmological constant $\Lambda$ has a new function in the mystical story and is driving the late-time unexpected expansion. To yet, however, it has not been well described in terms of the fine-tuning and cosmic coincidence conundrums~\cite{Copeland2006}. Therefore, the credit for starting the modelling of the accelerating Universe goes to Caldwell~\cite{Caldwell2002}, who was the first to characterise the dynamical nature of exotic energy by using a convenient equation of state (EOS) parameter ($\omega^{de}$). Later, Copeland~\cite{Copeland2006} introduced scalar field related dark energy concepts, viz. phantom, quintessence, tachyon-type models, etc. and examined the empirical evidence for the late-time accelerated expansion of the cosmic spacetime. Dynamical dark energy with a changeable EOS parameter is explored in specific examples for both spatially homogeneous and anisotropic spacetime~\cite{Kumar:2010vpj,Yadav:2011bj,Mishra:2017gtk,Mishra:2015jja,Ali:2015mov,Goswami:2016ppx,Yadav:2016zvc}. A few critical investigations on $\Lambda$-dominated era are available in the following Refs.~\cite{Zaeem-ul-HaqBhatti:2016moh,Yousaf:2017}. 

Some accurate analytical collapse solutions under the influence of in the absence of shear have been studied by several authors~\cite{Zaeem-ul-HaqBhatti:2016moh,Yousaf:2017}. Since then, the investigation of the impact of a specific condition on the formulation of an exact analytical solution of the relativistic stellar interior has been carried out under the assumption of (i) an expansion-free, (ii) non-static and (iii) non-diagonal cosmic filament filled with non-isotropic fluids~\cite{Yousaf:2017}. However, a recent modification to general relativity has focused cosmologists' attention on the cosmological constant concerns, leading them to find out the true origin of the late-time acceleration without turning to an exotic energy or cosmological constant. The $f(R,T)$ theory of gravity, suggested by Harko et al.~\cite{Harko2011}, is responsible for the shift in the geometrical part of the Einstein-Hilbert action. In this theory, the trace $T$ of the energy-momentum tensor and the Ricci scalar are vital to the matter Lagrangian. Consideration of the quantum field effect and the potential of the creation of particle is also at the forefront of $f(R,T)$ gravity. Astrophysical investigations would benefit greatly from considering such possibilities since they imply the existence of a link between the quantum theoretical concept and $f(R,T)$ gravity~\cite{Harko2011}. In the following references~\cite{Zubair:2016,Yousaf:2017,Moraes:2016,Das:2016,Singh:2015,Myrzakulov:2012,Houndjo:2011,Das2016,Das2017,Deb2018a,Deb2018b,Biswas2019,Deb2019a,Deb2019b,Biswas2020,Hulke2020,Maurya2020a,Maurya2020b,Maurya2020c,Biswas2021,Maurya2021,Maurya2023} one may have look into the significant uses of the $f(R, T)$ theory of gravity in the fields of astrophysics and cosmology. Interestingly, Ricci scalar ($R$) and trace of the energy-momentum tensor ($T$) are supposed to be linked with the matter Lagrangian in the $f(R,T)$ theory of gravity~\cite{Moraes2017,Sharma:2018ikm,Sharma:2019hqe}.

To describe the present destiny of the Universe, the typical FLRW cosmological models basically need an isotropic as well as homogeneous distribution of matter-energy component within the physical system. However, it has been argued that a spherically symmetric and spatially homogeneous cosmological model is essential to explain adequately the observed phenomena~\cite{Yadav:2010ge}. Therefore, with an isotropic and homogeneous fluid distribution in $f(R,T)$ gravity, the present work attempts to investigate a plausible model of the Universe within the observational constraints. 

Under the motivation mentioned above, we have outlined the manuscript as follows. In Sect.~\ref{sec2}, we provide a brief mathematical background of the metric and $f(R,T)$ Gravity. In Sect.~\ref{sec3}, the solution of the field equations using the power law of our current model is considered in detail. In Section~\ref{sec4}, we have done the observational analysis, i.e., a discussion of the observational constraints on the model parameters is provided. Physical parameters and diagnostic analysis have been done with the help of graphical plots in Sect.~\ref{sec5}. Finally, Sect.~\ref{sec6} is assigned to make some comments as the conclusion of the study.

\section{The metric and $f(R,T)$ Gravity} \label{sec2}
The usual form of the FRLW metric is given by
\begin{equation}
\label{ref41}
ds^{2}=-c^{2}dt^{2}+a^{2}\left(dx^{2}+dy^{2}+dz^{2}\right),
\end{equation}
where $a$ represents the scale factor, which is a function of $t$ alone.

The Einstein-Hilbert action under $f(R,T)$ gravity can be provided as
\begin{equation}
\label{basic}
S = \int{d^{4}x\sqrt{-g}L_{m}} + \frac{1}{16\pi}\int{d^{4}x\sqrt{-g}f(R,T)} ,
\end{equation}
where $g$ and $L_{m}$ are, respectively, the metric determinant and matter-Lagrangian density.

Then Einstein's general relativistic field equation in the $f(R,T)$ gravity is as follows:
\[
[f_{1}^{\prime}(R)+f_{2}^{\prime}(R)f_{3}^{\prime}(T)]R_{ij}-\frac{1}{2}f_{1}^{\prime}(R)g_{ij}+
\]
\[
(g_{ij}\nabla^{i}\nabla_{i}-\nabla_{i}\nabla_{j})[f_{1}^{\prime}(R)+f_{2}^{\prime}(R)f_{3}^{\prime}(T)]=
\]
\begin{equation}
\label{basic41}
[f_{2}^{\prime}(R)f_{3}^{\prime}(T)]T_{ij}+f_{2}(R)\left[f_{3}^{\prime}(T)p+\frac{1}{2}f_{3}(T)\right]g_{ij} + 8\pi,
\end{equation}
where $f(R,T)$ is assumed to have the form $f(R,T) = f_{1}(R)+f_{2}(R)f_{3}(T)$. In this case, the primes signify derivatives concerning the configuration in question. Additionally, we consider the cases with $f_{1}(R) = f_{2}(R) = R$ and $f_{3}(T) = \xi T$~\cite{Moraes2017}, where $\xi$ is a constant. 

Hence, Eq. (\ref{basic41}) in its formal structure can be written as 
\begin{equation}
\label{basic42}
G_{ij} = 8\pi T^{eff}_{ij} = 8\pi(T_{ij}+T^{ME}_{ij}),
\end{equation}
where the curvature tensor, the effective energy-momentum tensor, the matter energy-momentum tensor, and an additional energy term that is known to arise from the trace of the energy-momentum tensor $T$ are, respectively, $G_{ij}$, $T^{eff}_{ij}$, $T_{ij}$, and $T^{ME}_{ij}$~\cite{Bhardwaj2019}. The latter can be articulated as 

\begin{equation}
\label{basic3}
T^{ME}_{ij}=\frac{\xi R}{8\pi}\left(T_{ij}+\frac{3\rho-7p}{2}g_{ij}\right),
\end{equation}
where $p$ and $\rho$ are the pressure and the energy density of the perfect fluid of the cosmic structure.

Now, after application of the Bianchi identities in Eq. (\ref{basic42}), one can obtain
\begin{equation}
\label{basic4}
\nabla^{i}T_{ij} = -\frac{\xi R}{8\pi}\left[\frac{1}{2}g_{ij}\nabla^{i}(\rho-3p) + \nabla^{i}(T_{ij}+pg_{ij})\right].
\end{equation}

Hence, plugging Eq. (\ref{basic4}) in Eq. (\ref{basic41}), we can get
\begin{equation}
\label{basic5}
G_{ij}=\left(8\,\pi+\xi\right)\,T_{ij}-\left(1+\xi\right)\,R_{ij}+\frac{R}{2}\,\Big[1+\xi\,\left(2\,p+T\right)\Big]\,g_{ij},
\end{equation}
where $T_{ij}=\,\left(\rho+p\right)\,u_i\,u_j-p\,g_{ij}$, $c\,=\,1$, $u^i\,=\,(1,0,0,0)\,$, $\,u_i\,=\,(-1,0,0,0)\,$, $\,g_{ij}\,u^i\,u^j\,=\,-1$, $\frac{R}{2}\,=-\,3\,\left(\frac{\dot{a}^2}{a^2}+\frac{\ddot{a}}{a}\right),$ $T_{00}\,=\,2\,p+\rho\,$, $\,T_{ii}\,=\,-a^2\,p\,$, $\,i=1,2,3$ and $T\,=\,3\,p+\rho.$ Here $\dot{a}$ denotes a differentiation of the scale factor ($a$) with respect to the time coordinate ($t$).

Now, by solving Eqs. (\ref{basic5}) and (\ref{ref41}), we get
\begin{equation}
\label{basic5-14}
3H^{2} = 8\pi \left[\rho-\frac{3 \xi}{8\pi}(3\rho-7p)(\dot{H}+2H^{2})\right],
\end{equation}

\begin{equation}
\label{basic5-24}
2\dot{H}+3H^{2} = -8\pi\left[p+\frac{9\xi}{8\pi}(\rho-3p)(\dot{H}+2H^{2})\right],
\end{equation}
where $H (= \frac{\dot{a}}{a})$ is the Hubble parameter.

It is to be noted from Eqs. (\ref{basic5-14}) and (\ref{basic5-24}) that we have a physical system where there are two equations involved therein three unknown variables, viz. $H$, $\rho$ and $p$. As a result, we can not solve these equations in a general and straightforward way. Therefore, we must assume a parameterization scheme to find an explicit solution for $\rho$ and $p$. However, this scheme should be based on some basic requirements, including theoretically consistent and observational verification. 

For this purpose, in the next step, we shall employ the power law expansion~\cite{Kumar2012}, which can be shown to be appropriate to explore features of the observable Universe.

\section{Solution of the field equations by using power law} \label{sec3}
In the FRLW metric (\ref{ref41}), there is an unpredictable time-dependent function, denoted by $a(t)$. We can write $a(t) = \alpha e^{\beta t}$ with $\alpha$ and $\beta$. The acceleration of the cosmos at later times is analogous to the power law cosmology described by this form of $a(t)$~\cite{Kumar:2010si,Yadav:2010dj,Yadav:2010ah,Yadav:2011dr,Kumar2012,Sharma:2018ikm}.

The space-time (\ref{ref41}), which can be interpreted as follows
\begin{equation}
\label{ref1-2}
ds^{2}\,=\,-dt^{2}+\alpha^{2}t^{2\beta}\,\left(dx^{2}+dy^{2}+dz^{2}\right),
\end{equation}
where $\alpha$ and $\beta$ are arbitrary constants.

The deceleration parameter is given by~\cite{dec1,rayc1} 
\begin{equation}\label{uu1-12a-1}
q = -\frac{a{\ddot{a}}}{\dot{a}^{2}}=-\frac{\beta - 1}{\beta}.
\end{equation}

The scale factor is given by
\begin{equation}
\label{sf1}
a = \alpha t^{\beta},
\end{equation}
whereas the following expressions are obtained for the Hubble parameter, the scalar expansion and the proper volume, respectively~\cite{dec1,rayc1}:
\begin{equation}
\label{hp}
H =\frac{\beta}{t},
\end{equation}

\begin{equation}  \label{u215-c}
\Theta = 3H ,
\end{equation}

\begin{equation}  \label{u215-c}
\Theta  = \frac{3\ \beta}{t},
\end{equation}

\begin{equation}  \label{u218}
V\,=\,a^{3} =\; \alpha^{3}t^{3\beta}.
\end{equation}

In this connection, the shear tensor can be provided as
\begin{equation}  \label{u219}
\begin{array}{ll}
\sigma_{ij}\,=\,u_{(i;j)}-\frac{1}{3}\,\Theta\,(g_{ij}+u_i\,u_j) +\dot{u}_{(i}\,u_{j)},
\end{array}
\end{equation}
and the non-vanishing components of the $\sigma_i^j$ are
\begin{equation}  \label{u220}
\left\{
\begin{array}{ll}
\sigma_1^1\,=\,-\frac{2\,\beta}{t},\\
\\
\sigma_2^2\,=\,\sigma_3^3\,=\,\sigma_4^4\,=\,0.
\end{array}
\right.
\end{equation}

The expressions for the pressure $p$ and the energy density $\rho$, now from the Einstein field equations (\ref{basic5-14}) and (\ref{basic5-24}) can be obtained as follows:
\begin{equation}
\label{p}
p = -\frac{\beta(3\beta-2)\left[8\pi t^{2}-9\xi\beta(2\beta-1)\right] + 27\xi\beta^{2}(2\beta-1)}{\left[64\pi^{2}t^{2}-288\pi\xi\beta(2\beta-1)\right]t^{2} + 54\xi^{2}\beta^{2}(2\beta-1)^{2}},
\end{equation}

\begin{equation}
\label{rho}
\rho = \frac{3\beta^{2}[8\pi t^{2}-27\xi\beta(2\beta-1)] + 21\xi\beta^{2}(3\beta-2)(2\beta-1)}{\left[64\pi^{2}t^{2}-288\pi\xi\beta(2\beta-1)\right]t^{2}+54\xi^{2}\beta^{2}(2\beta-1)^{2}}.
\end{equation}

It is to be noted that our model can easily retrieve the scenario of general relativity (GR) for the imposed condition $\xi = 0$. As a result, the expressions of the cosmic pressure and density can be recovered as follows:
\begin{equation}
\label{p-1}
p=\frac{\beta(3\beta-2)}{8\pi t^{2}},
\end{equation} 

\begin{equation}
\label{rho-1}
\rho = \frac{3\beta^{2}}{8\pi t^{2}}.
\end{equation}

\section{Observational Analysis: Constraints on Model Parameters} \label{sec4}
In this Section, we try to limit the model parameters $H_o$, $\alpha$ and $\beta$ with reference to the observationally obtained $H(z)$ dataset under the redshift range $0 \leq z \leq 1.965$. The $H(z)$ observational dataset is provided in the references. Moreover, the scaling factor with respect to redshift is given by
\begin{equation}
a = \frac{a_0}{z+1} = \alpha t^{\beta}, \label{eq: a equation}
\end{equation} 
where $a_0$ denotes the present value of the scale factor.

The age of the Universe is computed with the following equation
\begin{equation}
    H(z) = - \frac{1}{z+1} \frac{dz}{dt}.
    \label{eq: H diff eqn}
\end{equation}

Putting together Eqs. (\ref{eq: a equation} -- \ref{eq: H diff eqn}) and there after a bit manipulation, we can easily get the following Hubble parameter in its functional form as
\begin{equation} 
H(z) = \beta \left(\frac{a_0}{\alpha}\right)^{-\frac{1}{\beta}} (z+1)^{\frac{1}{\beta}}
    \label{eq: model equation}
\end{equation}

From Eq. (\ref{eq: model equation}), the present value of the Hubble constant is determined as $H_0 = \beta \left(\frac{a_0}{\alpha}\right)^{-\frac{1}{\beta}}$.

\begin{figure}
  \centering
  \includegraphics[width=0.50\textwidth]{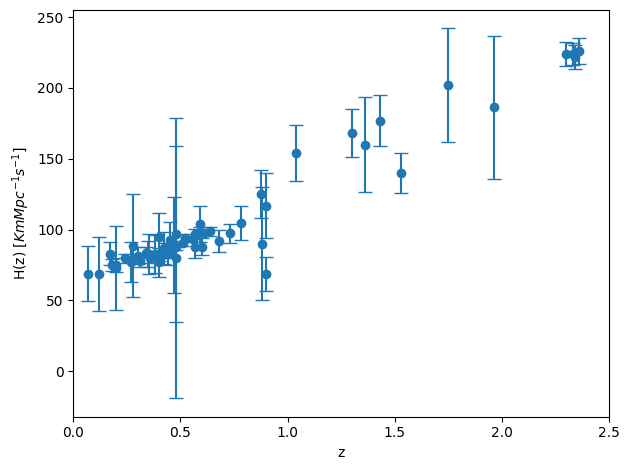}
  \caption{Error bar plot of the 57 point $H(z)$ data used in analysis of our model}
  \label{fig:H(z) data}
\end{figure}

Let us now confine the model parameters, viz. $H_0$, $\alpha$ and $\beta$ to the redshift range $ 0 \leq z \leq 1.965$ by using the observable $H(z)$, BAO and Pantheon datasets. The distribution of the datasets for the observational aspect of $H(z)$ is shown in \ref{fig:H(z) data} and the data points are listed in \ref{tab:57datapt}. The details for BAO and Pantheon are taken from ref.~\cite{Akarsu2019}.

%%%%%%%%%%%%%%%%%%%%%%%%%%%%%%%%%%%%%%%%%%%%%%%%%%%%%%%%%%%%%%%%%%%%%%%%%%%%%%%%%%%%%%%%%%%%%%%%%%%%%%%%%%%%%%%%%%%%%%%%%%%%%%%%%%%%%%%%%%%%%%%%%
\begin{table*}
\caption{Parameter values obtained from different datasets after running MCMC and Bayesian analysis where $t_C$ is the age of the Universe (Gyr) with the notations as $H(z)_1 = H(z) + Pantheon$ and $H(z)_2 = H(z) + Pantheon + BAO$}
\centering
\label{Parameter Values Table}
\renewcommand{\arraystretch}{1.5}
\setlength{\tabcolsep}{11pt}
% \footnotesize
\begin{tabular}{lccccc}
\hline
\textbf{Parameter} & \textbf{$H(z)$} & \textbf{BAO} & \textbf{Pantheon} & \textbf{$H(z)_1$} & \textbf{$H(z)_2$} \\
\hline
$H_0$  & $67.098^{+2.148}_{-1.792}$ & $67.588^{+2.229}_{-2.170}$ & $66.270^{+2.215}_{-2.181}$ & $65.960^{+2.380}_{-1.834}$ & $66.274^{+2.015}_{-1.864}$\\
$\alpha$ & $67.014^{+0.096}_{-0.087}$ & $67.016^{+0.084}_{-0.111}$ & $66.998^{+0.090}_{-0.092}$ & $66.998^{+0.099}_{-0.107}$ & $67.030^{+0.114}_{-0.126}$ \\
$\beta$ & $1.000^{+0.008}_{-0.010}$ & $0.997^{+0.011}_{-0.010}$ & $1.003^{+0.011}_{-0.010}$ & $1.005^{+0.009}_{-0.011}$ & $1.004^{+0.009}_{-0.009}$ \\
$t_C$ & $14.903574$ & $14.7511392$ & $15.1350535$ & $15.2365069$ & $15.1492289$ \\

\hline
\end{tabular}
\end{table*}
%%%%%%%%%%%%%%%%%%%%%%%%%%%%%%%%%%%%%%%%%%%%%%%%%%%%%%%%%%%%%%%%%%%%%%%%%%%%%%%%%%%%%%%%%%%%%%%%%%%%%%%%%%%%%%%%%%%%%%%%%%%%%%%%%%%%%%%%%%%%%%%%%%

%%%%%%%%%%%%%%%%%%%%%%%%%%%%%%%%%%%%%%%%%%%%%%%%%%%%%%%%%%%%%%%%%%%%%%%%%%%%%%%%%%
%%%%%%%%%%%%%%%%%%%%%%%%%%%%%%%%%%%%%%%%%%%%%%%%%%%%%%%%%%%%%%%%%%%%%%%%%%%%%%%%%%

\newpage

%%%%%%%%%%%%%%%%%%%%%%%%%%%%%%%%%%%%%%%%%%%%%%%%%%%%%%%%%%%%%%%%%%%%%%%%%%%%%%%%%%
\begin{figure}
\includegraphics[width=.50\textwidth]{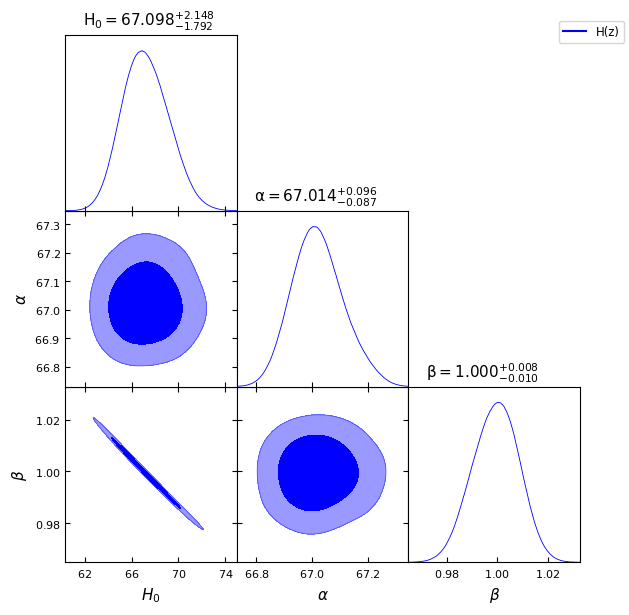}
\caption{One-dimensional marginalized distribution and two-dimensional contours for our $f(R,T)$ model parameters $H_0$, $\alpha$, $\beta$ using the $H(z)$ dataset presented in Table \ref{tab:57datapt}} \label{H(z)}
\end{figure}
%%%%%%%%%%%%%%%%%%%%%%%%%%%%%%%%%%%%%%%%%%%%%%%%%%%%%%%%%%%%%%%%%%%%%%%%%%%%%%%%%%

%%%%%%%%%%%%%%%%%%%%%%%%%%%%%%%%%%%%%%%%%%%%%%%%%%%%%%%%%%%%%%%%%%%%%%%%%%%%%%%%%%
\begin{figure}
\includegraphics[width=.50\textwidth]{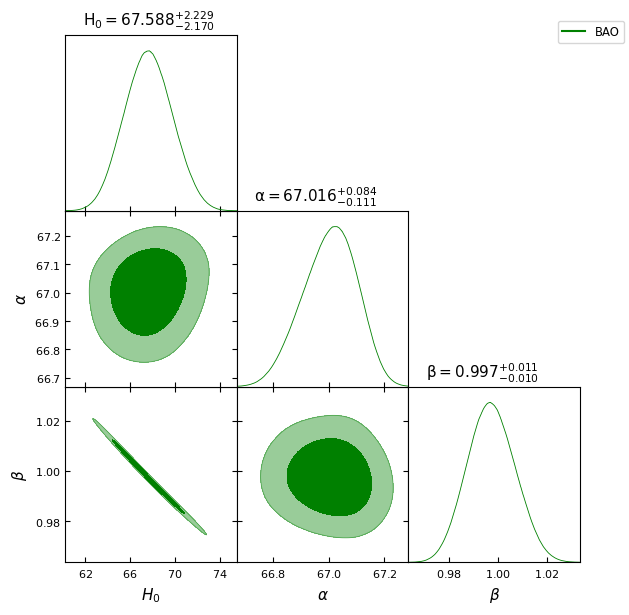}
\caption{One-dimensional marginalized distribution and two-dimensional contours for our $f(R,T)$ model parameters $H_0$, $\alpha$ and $\beta$ using the BAO dataset ref.~\cite{Akarsu2019}} \label{BAO}
\end{figure}
%%%%%%%%%%%%%%%%%%%%%%%%%%%%%%%%%%%%%%%%%%%%%%%%%%%%%%%%%%%%%%%%%%%%%%%%%%%%%%%%%%

%%%%%%%%%%%%%%%%%%%%%%%%%%%%%%%%%%%%%%%%%%%%%%%%%%%%%%%%%%%%%%%%%%%%%%%%%%%%%%%%%%
\begin{figure}
\includegraphics[width=.50\textwidth]{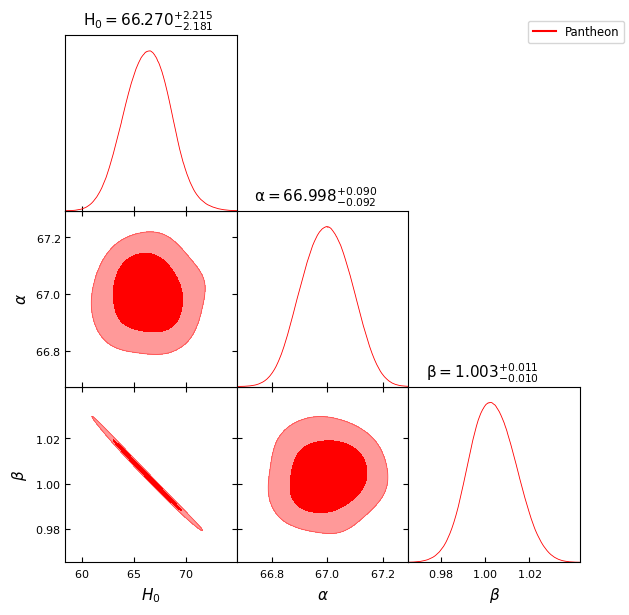}
\caption{One-dimensional marginalized distribution and two-dimensional contours for our $f(R,T)$ model parameters $H_0$, $\alpha$ and $\beta$ using the Pantheon dataset ref.~\cite{Akarsu2019}} \label{Pantheon}
\end{figure}
%%%%%%%%%%%%%%%%%%%%%%%%%%%%%%%%%%%%%%%%%%%%%%%%%%%%%%%%%%%%%%%%%%%%%%%%%%%%%%%%%%

%%%%%%%%%%%%%%%%%%%%%%%%%%%%%%%%%%%%%%%%%%%%%%%%%%%%%%%%%%%%%%%%%%%%%%%%%%%%%%%%%%
\begin{figure}
\includegraphics[width=.50\textwidth]{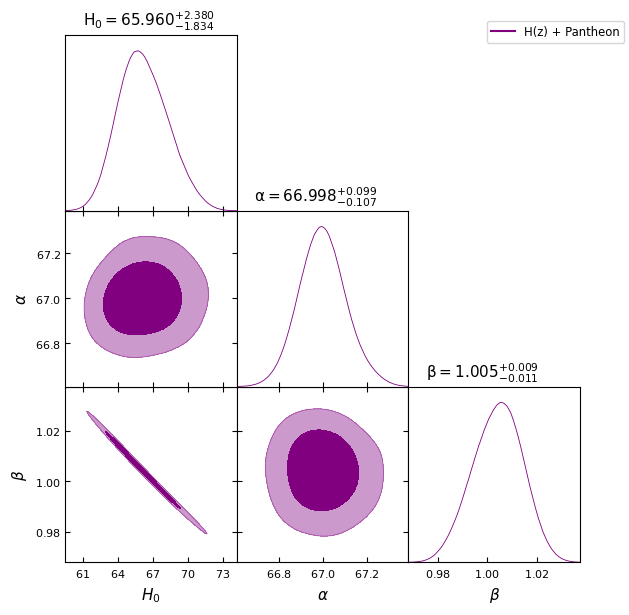}
\caption{One-dimensional marginalized distribution and two-dimensional contours for our $f(R,T)$ model parameters $H_0$, $\alpha$ and $\beta$ using the combination of $H(z)$ and Pantheon dataset} \label{H(z) + Pantheon}
\end{figure}
%%%%%%%%%%%%%%%%%%%%%%%%%%%%%%%%%%%%%%%%%%%%%%%%%%%%%%%%%%%%%%%%%%%%%%%%%%%%%%%%%%

%%%%%%%%%%%%%%%%%%%%%%%%%%%%%%%%%%%%%%%%%%%%%%%%%%%%%%%%%%%%%%%%%%%%%%%%%%%%%%%%%%
\begin{figure}
\includegraphics[width=.50\textwidth]{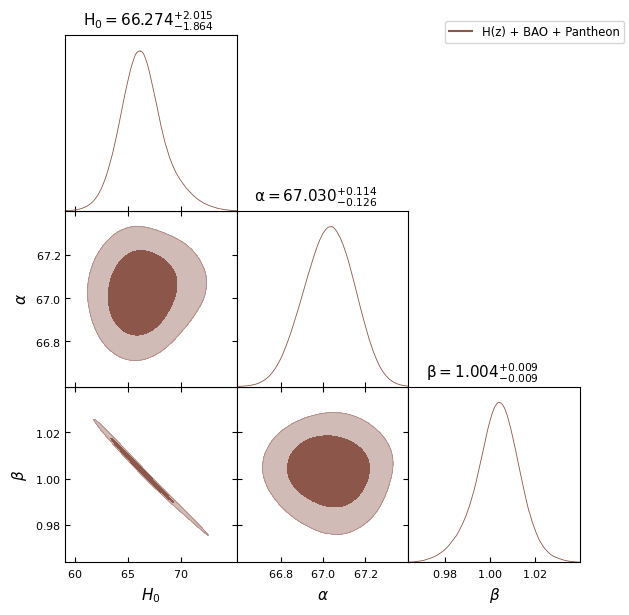}
\caption{One-dimensional marginalized distribution and two-dimensional contours for our $f(R,T)$ model parameters $H_0$, $\alpha$ and $\beta$ using the combination of $H(z)$, BAO and Pantheon dataset} \label{H(z) + BAO + Pantheon}
\end{figure}
%%%%%%%%%%%%%%%%%%%%%%%%%%%%%%%%%%%%%%%%%%%%%%%%%%%%%%%%%%%%%%%%%%%%%%%%%%%%%%%%%%

%%%%%%%%%%%%%%%%%%%%%%%%%%%%%%%%%%%%%%%%%%%%%%%%%%%%%%%%%%%%%%%%%%%%%%%%%%%%%%%%%%
\begin{figure}
\includegraphics[width=.50\textwidth]{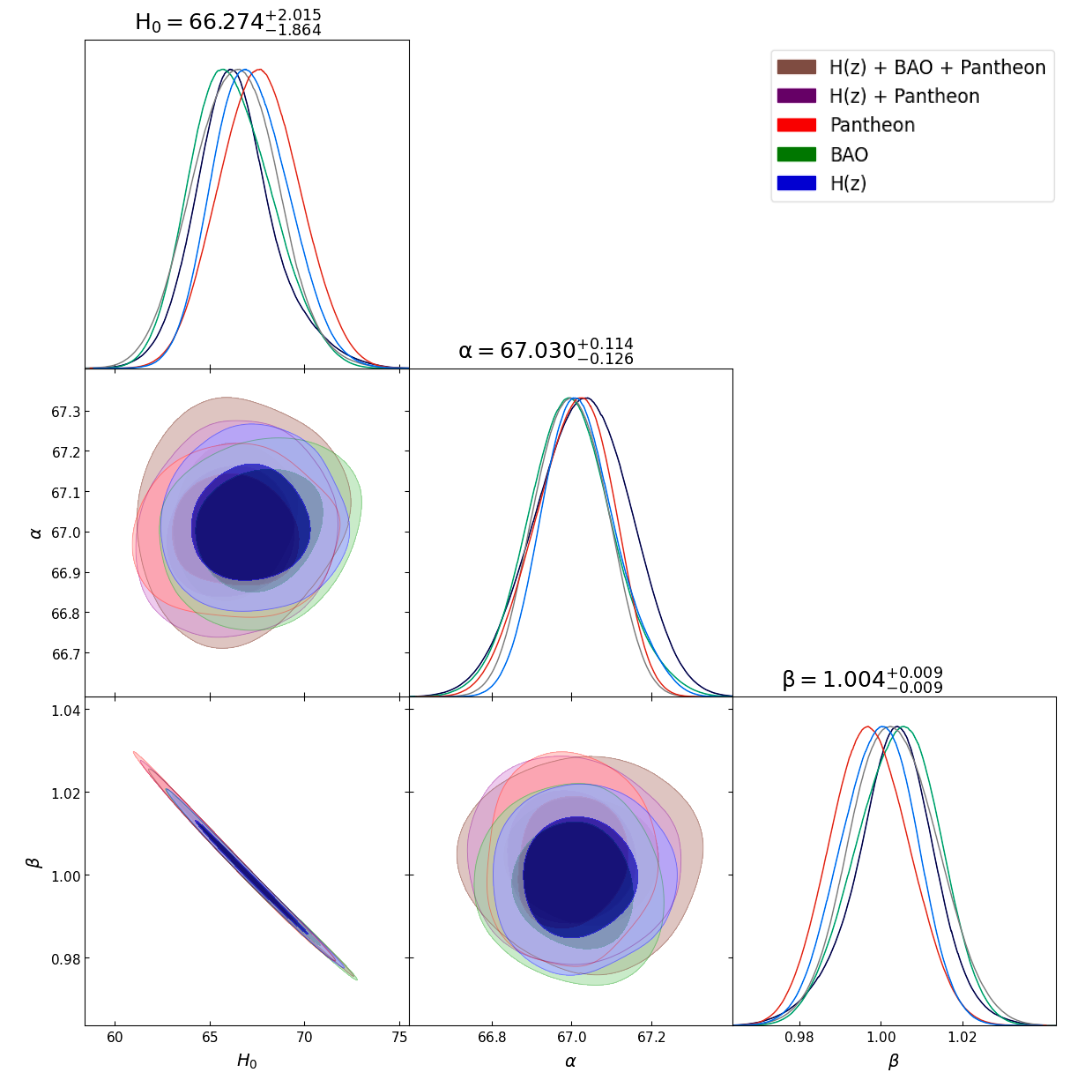}
\caption{One-dimensional marginalized distributions and two-dimensional contours representing our $f(R,T)$ model parameters $H_0$, $\alpha$ and $\beta$, depicting the combined variability across all dataset combinations} \label{All Combined}
\end{figure}
%%%%%%%%%%%%%%%%%%%%%%%%%%%%%%%%%%%%%%%%%%%%%%%%%%%%%%%%%%%%%%%%%%%%%%%%%%%%%%%%%%

%%%%%%%%%%%%%%%%%%%%%%%%%%%%%%%%%%%%%%%%%%%%%%%%%%%%%%%%%%%%%%%%%%%%%%%%%%%%%%%%%%
\begin{figure}[htbp]
  \centering
  %\begin{subfigure}{0.50\textwidth}
    \includegraphics[width=0.35\textwidth]{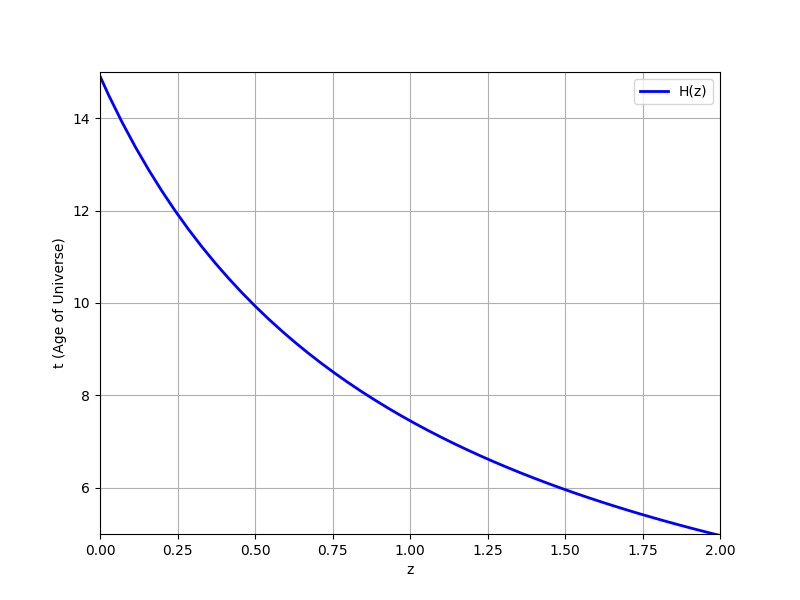}
    %\caption{}
  %\end{subfigure}
  %\begin{subfigure}{0.50\textwidth}
    \includegraphics[width=0.35\textwidth]{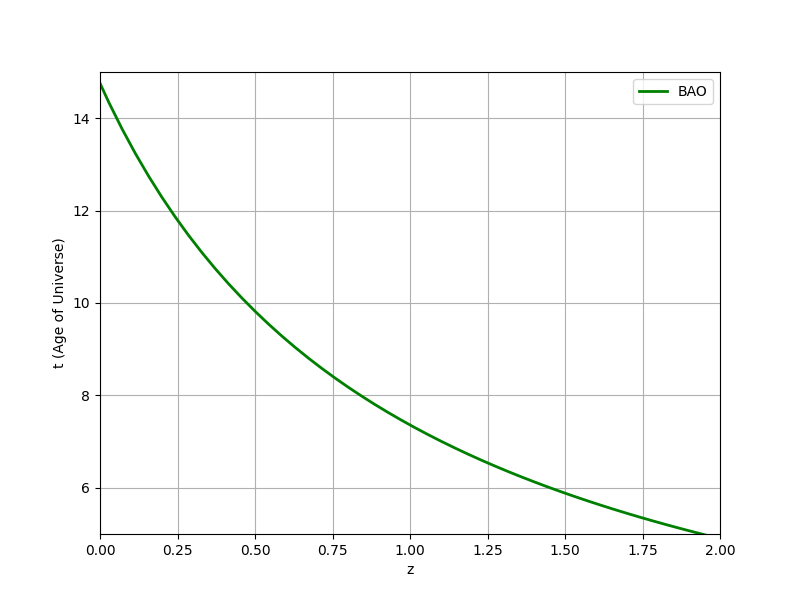}
    %\caption{}
  %\end{subfigure}
  %\begin{subfigure}{0.50\textwidth}
    \includegraphics[width=0.35\textwidth]{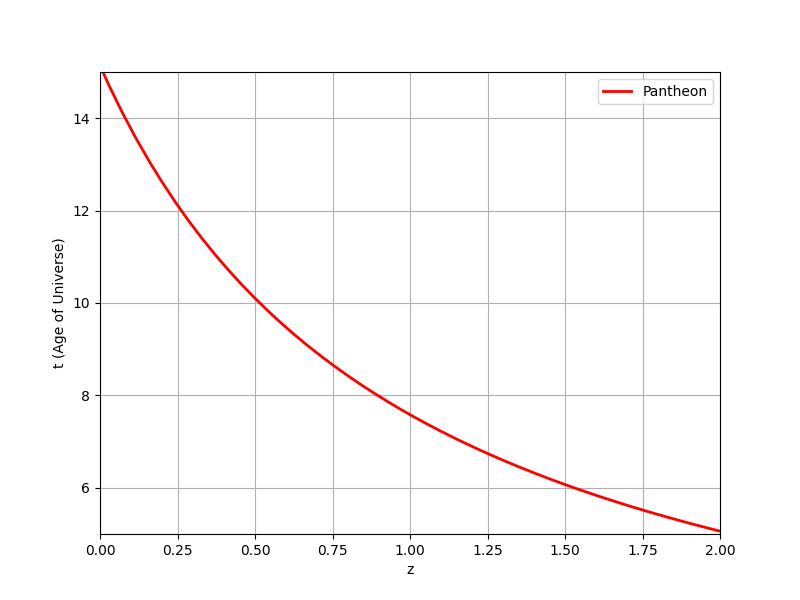}
    %\caption{}
  %\end{subfigure}
  %\begin{subfigure}{0.50\textwidth}
    \includegraphics[width=0.35\textwidth]{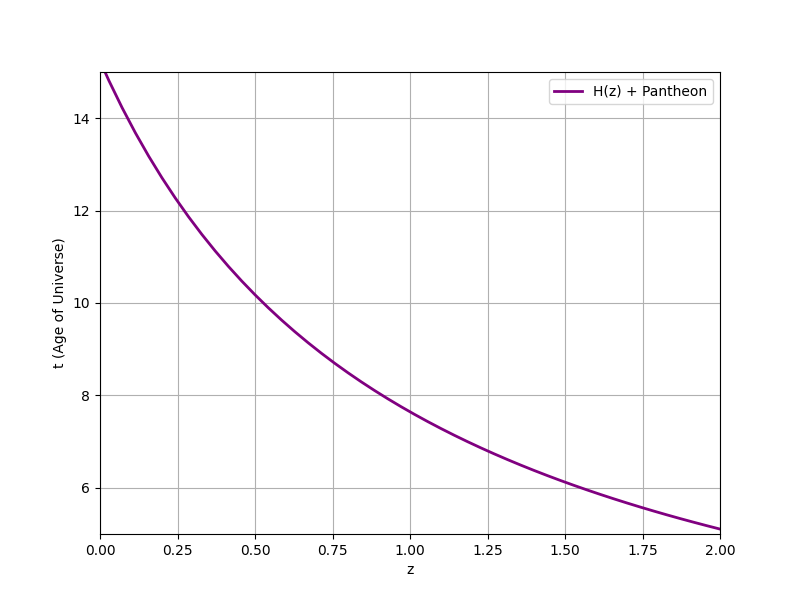}
    %\caption{}
  %\end{subfigure}
  %\begin{subfigure}{0.50\textwidth}
    \includegraphics[width=0.35\textwidth]{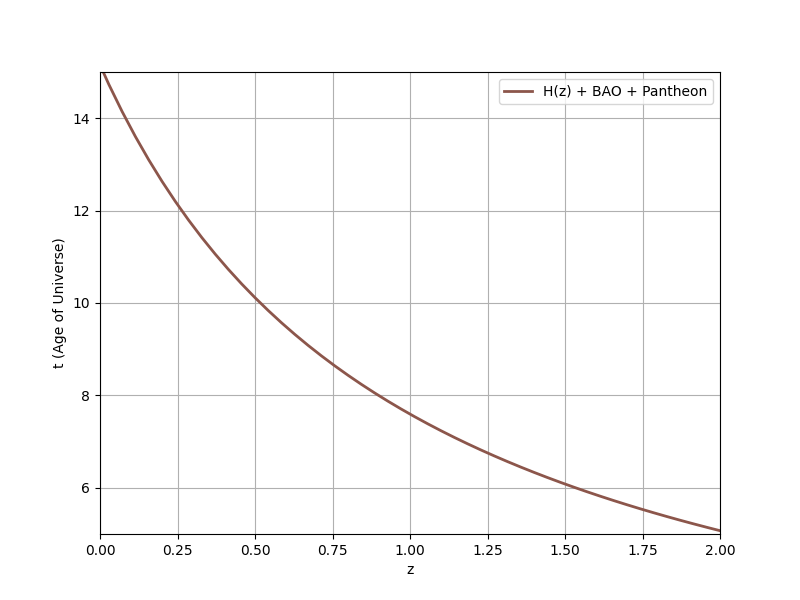}
    %\caption{}
  %\end{subfigure}
  \caption{Graphical representation of cosmic time over redshift. It is crucial for understanding cosmic evolution and validating cosmological models. Five plots are for different combinations of $H(z)$, BAO and Pantheon datasets. The age of the Universe ($t_C$) at $z = 0$ are mentioned in the Table \ref{Parameter Values Table}}
  \label{fig:statfind}
\end{figure}
%%%%%%%%%%%%%%%%%%%%%%%%%%%%%%%%%%%%%%%%%%%%%%%%%%%%%%%%%%%%%%%%%%%%%%%%%%%%%%%%%%

\section{Physical parameters and diagnostic analysis} \label{sec5}

\subsection{Energy Conditions}
Energy conditions (ECs) are like the cosmic rules that help us to understand that the universe (energy as well as matter) distributes throughout the universe. They are based on the Einstein equations of gravity and serve as the laws of the cosmos. These conditions reveal the distribution of the energy and matter in space.

\begin{enumerate}
    \item Weak Energy Condition (WEC): 
        The WEC says that the energy cannot be negative or zero anywhere in the universe. This condition helps us keep the universe's rules fair and consistent. Figure \ref{fig:WEC} shows the nature of WEC for our model using the parameters obtained in the Bayesian Analysis.
        
    \item Null Energy Condition (NEC): 
        The NEC is about light, which states that when light travels through the universe, it always encounters some energy. This condition keeps the universe's physics sensible and prevents strange things from happening. Figure \ref{fig:NEC} shows the nature of NEC for our model using the parameters obtained in the Bayesian Analysis.
        
    \item Strong Energy Condition (SEC): 
        The SEC is like a stricter version of the NEC. It ensures that energy cannot be negative and limits how things behave under gravity. It is like saying gravity always pulls things together; it cannot push them apart. This rule helps maintain order in the universe. Figure \ref{fig:SEC} shows the nature of SEC for our model using the parameters obtained in the Bayesian Analysis.
        
    \item Dominant Energy Condition (DEC): 
        The DEC builds on the NEC and ensures that energy is not only non-negative but that how energy flows around cannot be too wild. It is like saying that energy cannot move faster than light or get too crazy. This condition ensures that the universe does not have any weird surprises. Figure \ref{fig:DEC} shows the nature of DEC for our model using the parameters obtained in the Bayesian Analysis.
\end{enumerate}

\begin{figure}[htbp]
  \centering
  \includegraphics[width=0.50\textwidth]{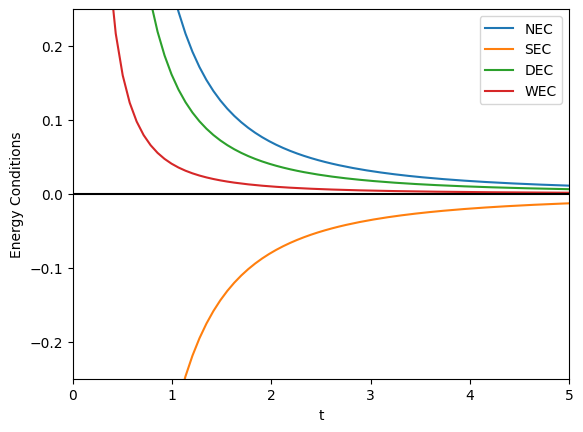}
  \caption{Visualisation of all Energy Conditions vs time}
  \label{fig:energy}
\end{figure}

All the above-mentioned energy conditions are combined in Fig. \ref{fig:energy} whereas the energy condition relations are as shown mathematically below~\cite{Singh2015,Singh2023}:
\begin{align}
 \text{NEC}: \rho + p \geq 0, \\
 \text{WEC}: \rho + p \geq 0~\&~\rho \geq 0, \\
 \text{SEC}: \rho + p \geq 0~\&~\rho + 3p \geq 0,\\
  \text{DEC}: \rho - p \geq 0~\&~\rho \geq 0. 
\end{align}

\begin{figure}[htbp]
  \centering
  \includegraphics[width=0.50\textwidth]{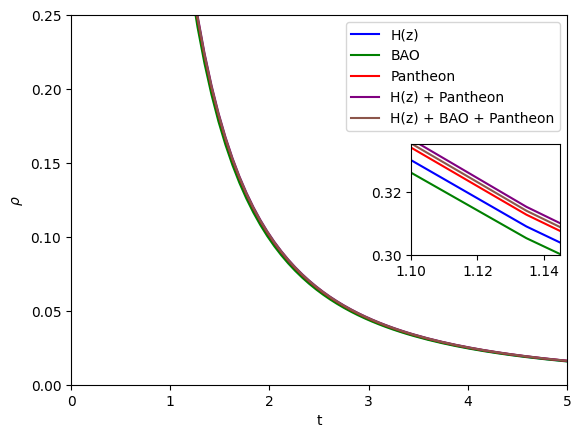}
  \caption{Graphical representation illustrates the dynamic variation of the energy density ($\rho$) over the time ($t$) under various parameter conditions derived from distinct combinations of the $H(z)$, BAO and Pantheon datasets}
  \label{fig:rho}
\end{figure}

\begin{figure}[htbp]
  \centering
  \includegraphics[width=0.50\textwidth]{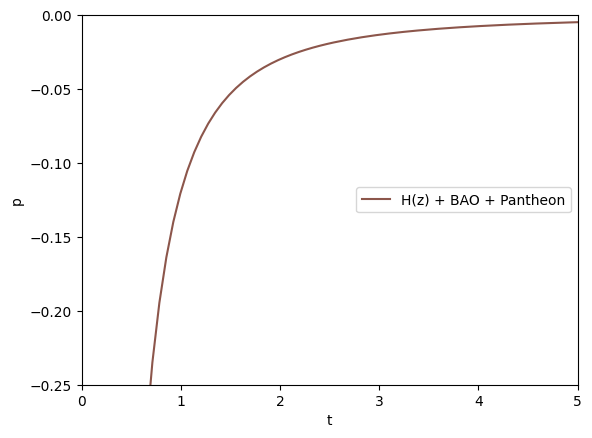}
  \caption{Graphical representation illustrates the dynamic variation of the pressure $p$ over time ($t$) under various parameter conditions derived from distinct combinations of the $H(z)$, BAO and Pantheon datasets}
  \label{fig:p}
\end{figure}

Fig. \ref{fig:rho} displays the energy density distribution $\rho$ with respect to time $t$, whereas Fig. \ref{fig:p} illustrates the pressure $p$. Our results demonstrate that all other energy conditions i.e., NEC, WEC, and DEC, are met, except the SEC. The universe's growing rate supports the SEC violation. Consequently, the late-time acceleration of the present universe may be satisfactorily explained by the $f(R,T)$ theory of gravity, which benefits from the trace energy $T$ contribution without requiring the presence of dark energy or the cosmological constant in the universe's energy content.

%%%%%%%%%%%%%%%%%%%%%%%%%%%%%%%%%%%%%%%%%%%%%%%%%%%%%%%%%%%%%%%%%%%%%%%%%%%%%%%%%%
%%%%%%%%%%%%%%%%%%%%%%%%%%%%%%%%%%%%%%%%%%%%%%%%%%%%%%%%%%%%%%%%%%%%%%%%%%%%%%%%%%

%%%%%%%%%%%%%%%%%%%%%%%%%%%%%%%%%%%%%%%%%%%%%%%%%%%%%%%%%%%%%%%%%%%%%%%%%%%%%%%%%%
%%%%%%%%%%%%%%%%%%%%%%%%%%%%%%%%%%%%%%%%%%%%%%%%%%%%%%%%%%%%%%%%%%%%%%%%%%%%%%%%%%

%%%%%%%%%%%%%%%%%%%%%%%%%%%%%%%%%%%%%%%%%%%%%%%%%%%%%%%%%%%%%%%%%%%%%%%%%%%%%%%%%%
%%%%%%%%%%%%%%%%%%%%%%%%%%%%%%%%%%%%%%%%%%%%%%%%%%%%%%%%%%%%%%%%%%%%%%%%%%%%%%%%%%

%%%%%%%%%%%%%%%%%%%%%%%%%%%%%%%%%%%%%%%%%%%%%%%%%%%%%%%%%%%%%%%%%%%%%%%%%%%%%%%%%%
\begin{figure}[htbp]
  \centering
  \includegraphics[width=0.50\textwidth]{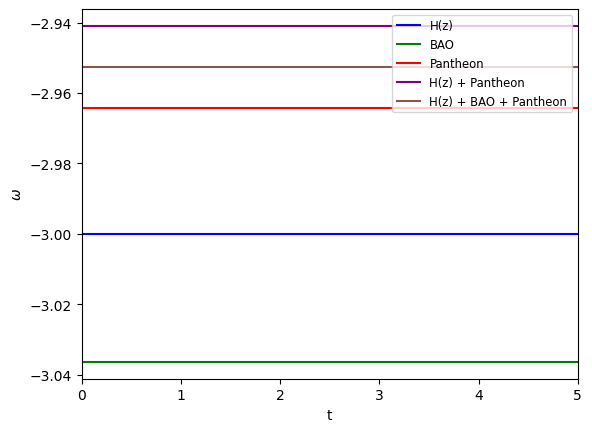}
  \caption{Plot of variation of Equation of State parameter $\omega$ vs time t The plot suggests that dark energy contributes to the universe's accelerated expansion but with some variations over time, potentially leading to interesting cosmological consequences}
  \label{fig:sep_energy}
\end{figure}
%%%%%%%%%%%%%%%%%%%%%%%%%%%%%%%%%%%%%%%%%%%%%%%%%%%%%%%%%%%%%%%%%%%%%%%%%%%%%%%%%%

%%%%%%%%%%%%%%%%%%%%%%%%%%%%%%%%%%%%%%%%%%%%%%%%%%%%%%%%%%%%%%%%%%%%%%%%%%%%%%%%%%
\begin{figure}[htbp]
  \centering
  \includegraphics[width=0.50\textwidth]{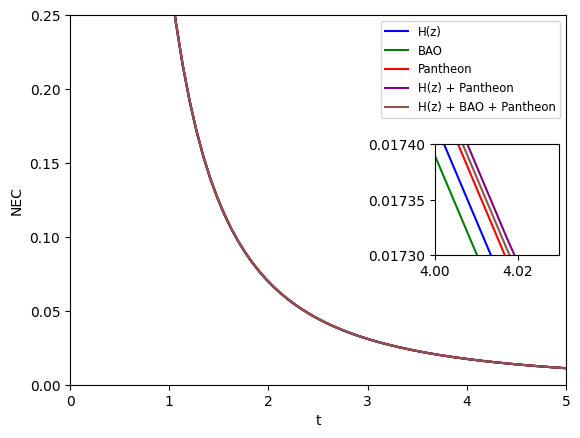}
  \caption{Visualisation of null energy condition (NEC) vs time for all dataset combinations}
  \label{fig:NEC}
\end{figure}
%%%%%%%%%%%%%%%%%%%%%%%%%%%%%%%%%%%%%%%%%%%%%%%%%%%%%%%%%%%%%%%%%%%%%%%%%%%%%%%%%%

%%%%%%%%%%%%%%%%%%%%%%%%%%%%%%%%%%%%%%%%%%%%%%%%%%%%%%%%%%%%%%%%%%%%%%%%%%%%%%%%%%
\begin{figure}[htbp]
  \centering
  \includegraphics[width=0.50\textwidth]{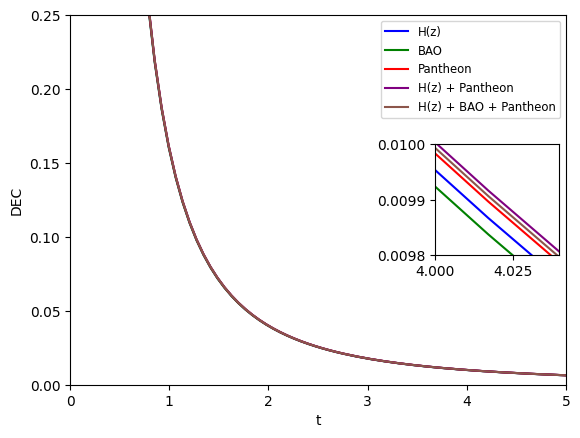}
  \caption{Visualisation of the dominant energy condition (DEC) vs time for all dataset combinations}
  \label{fig:DEC}
\end{figure}
%%%%%%%%%%%%%%%%%%%%%%%%%%%%%%%%%%%%%%%%%%%%%%%%%%%%%%%%%%%%%%%%%%%%%%%%%%%%%%%%%%

%%%%%%%%%%%%%%%%%%%%%%%%%%%%%%%%%%%%%%%%%%%%%%%%%%%%%%%%%%%%%%%%%%%%%%%%%%%%%%%%%%
\begin{figure}[htbp]
  \centering
  \includegraphics[width=0.50\textwidth]{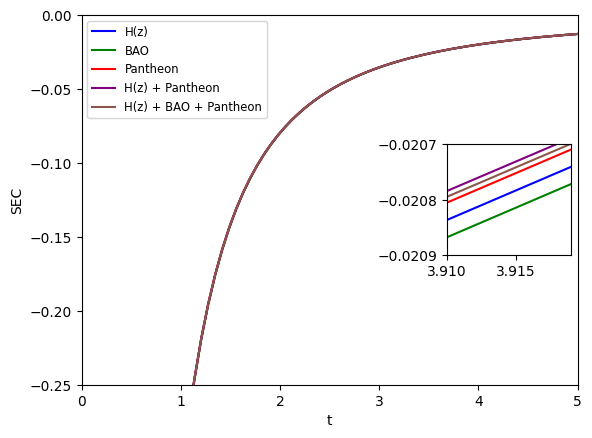}
  \caption{Visualisation of the strong energy conditions (SEC) vs time for all dataset combinations}
  \label{fig:SEC}
\end{figure}
%%%%%%%%%%%%%%%%%%%%%%%%%%%%%%%%%%%%%%%%%%%%%%%%%%%%%%%%%%%%%%%%%%%%%%%%%%%%%%%%%%

%%%%%%%%%%%%%%%%%%%%%%%%%%%%%%%%%%%%%%%%%%%%%%%%%%%%%%%%%%%%%%%%%%%%%%%%%%%%%%%%%%
\begin{figure}[htbp]
  \centering
  \includegraphics[width=0.50\textwidth]{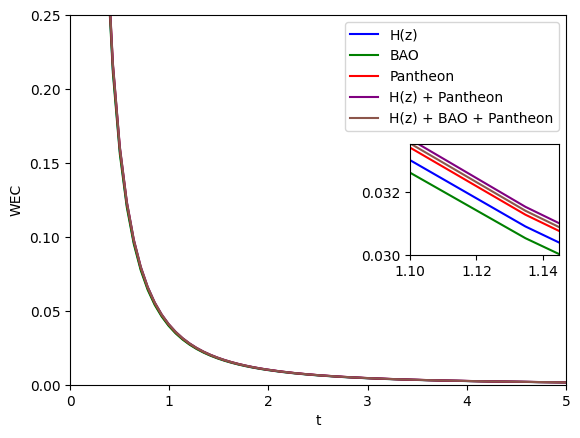}
  \caption{Visualisation of the weak energy conditions (WEC) vs time for all dataset combinations}
  \label{fig:WEC}
\end{figure}
%%%%%%%%%%%%%%%%%%%%%%%%%%%%%%%%%%%%%%%%%%%%%%%%%%%%%%%%%%%%%%%%%%%%%%%%%%%%%%%%%%

\subsection{State finder diagnostic}
State Finder Diagnostics are like cosmic detectives helping us unravel the mysteries of dark energy and the universe's evolution. These diagnostics are like a cosmic compass, guiding us through the complexities of cosmic evolution. State Finder Diagnostics build upon the two key parameters $r$ and $s$. These parameters help us understand how the universe is changing over time. One can think of them as the cosmic meters that tell us about the universe's expansion and the stuff that makes it up. They are dimensionless parameters that encapsulate the essence of the universe's solution, providing a lens through which we discern the universe's underlying dynamics. The general mathematical parametric definitions of these factors are as follows:
\begin{equation}
    r = \frac{\ddot{\dot{a}}}{aH^3}, \\
    r = \frac{(-1+r)}{3(-\frac{1}{2} + q)}. 
\end{equation}

However, the equations for $r$ and $s$ in our model which are expressed in terms of $q$, can be obtained as follows:
\begin{align*}
r &= q(1+2q), \\
s &= \frac{2}{3}(q+1).
\end{align*}

Figure \ref{fig:statfind} has demonstrated that the scale factor trajectories in the derived model follow a particular set of routes. Our approach corresponds to the outcomes obtained from the power law cosmology for the cosmic diagnostic pair.

%%%%%%%%%%%%%%%%%%%%%%%%%%%%%%%%%%%%%%%%%%%%%%%%%%%%%%%%%%%%%%%%%%%%%%%%%%%%%%%%%%
\begin{figure}[htbp]
  \centering
  %\begin{subfigure}{\textwidth}
    \includegraphics[width=0.45\textwidth]{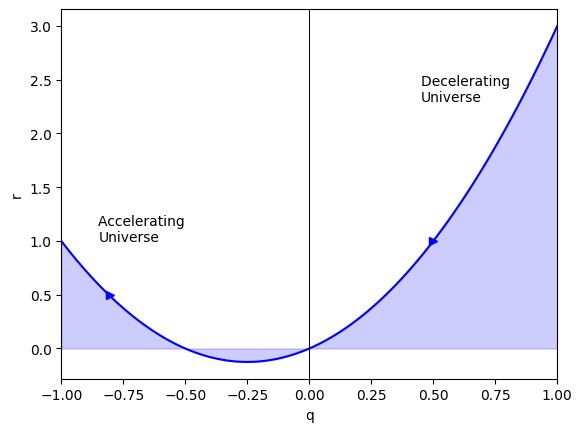}
    %\caption{}
  %\end{subfigure}
  %\begin{subfigure}{\textwidth}
    \includegraphics[width=0.45\textwidth]{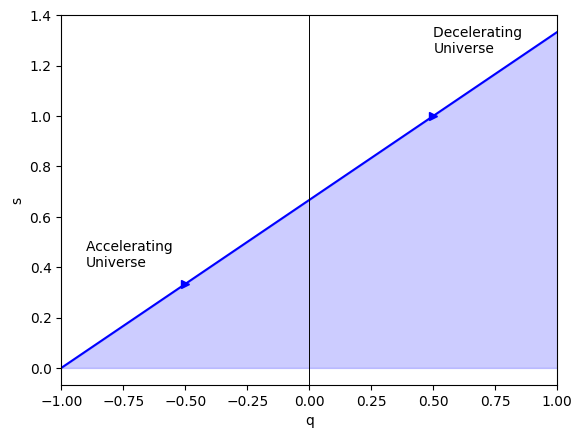}
    %\caption{}
 % \end{subfigure}
  %\begin{subfigure}{\textwidth}
    \includegraphics[width=0.45\textwidth]{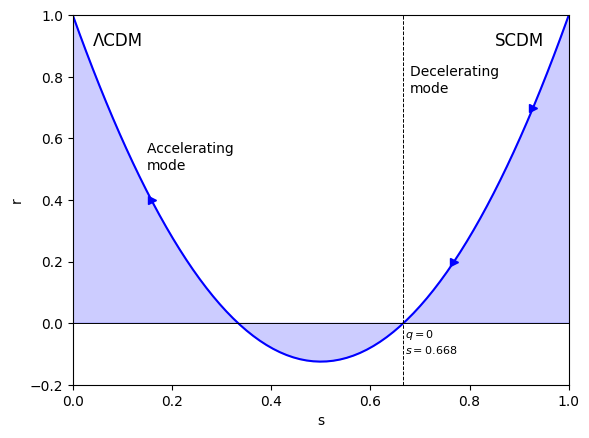}
    %\caption{}
  %\end{subfigure}
  \caption{State finder plots of $r-q$, $s-q$ and $r-q$.}
  \label{fig:statfind}
\end{figure}
%%%%%%%%%%%%%%%%%%%%%%%%%%%%%%%%%%%%%%%%%%%%%%%%%%%%%%%%%%%%%%%%%%%%%%%%%%%%%%%%%%

\subsection{Om(z) parameter}
Researchers typically use the state finder parameters $r - s$ and do analysis using the Om diagnostic to look at different dark energy theories in their academic works. The Hubble parameter H and the cosmic redshift z combine to generate the $Om(z)$ parameter, which is essential. Initially, Sahni et al.~\cite{Sahni2008} and later on others~\cite{Tripathy2021,Mandal2022,Myrzakulov2023} offer the $Om(z)$ parameter formula in the context of changed gravity as follows:
\begin{equation}
    Om(z) = \frac{[\frac{H(z)}{H_0}]^2 - 1}{(1+z)^3 - 1}.
\end{equation}

Here, the Hubble parameter is denoted by $H_0$. According to Shahalam et al.~\cite{Shahalam2015}, the quintessence ($\omega \ge -1$), $\Lambda$CDM and phantom ($\omega \le -1$) dark energy (DE) models are represented by the negative, zero and positive values of $Om(z)$, respectively. We obtain the $Om(z)$ parameter for the present model as follows:
\begin{equation}
    Om(z) = \frac{(1+z)^{2/b}-1}{(1+z)^3-1}.
\end{equation}

%%%%%%%%%%%%%%%%%%%%%%%%%%%%%%%%%%%%%%%%%%%%%%%%%%%%%%%%%%%%%%%%%%%%%%%%%%%%%%%%%%
\begin{figure}[htbp]
  \centering
  \includegraphics[width=.50\textwidth]{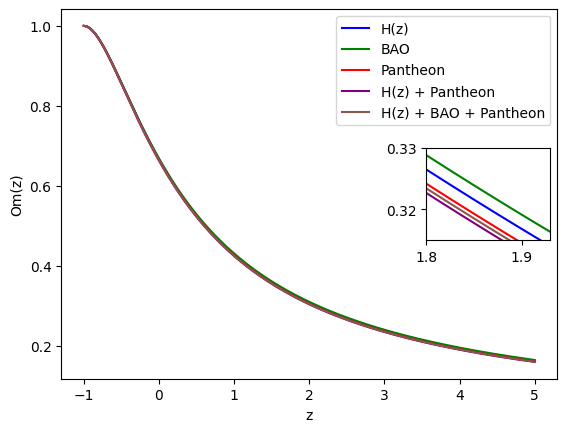}
  \caption{Examining the variations of $Om(z)$ with $z$ across different dataset combinations, considering the $\beta$ values obtained from each dataset.}
  \label{fig:om(z)}
\end{figure}
%%%%%%%%%%%%%%%%%%%%%%%%%%%%%%%%%%%%%%%%%%%%%%%%%%%%%%%%%%%%%%%%%%%%%%%%%%%%%%%%%%

%%%%%%%%%%%%%%%%%%%%%%%%%%%%%%%%%%%%%%%%%%%%%%%%%%%%%%%%%%%%%%%%%%%%%%%%%%%%%%%%%%
\begin{figure}[htbp]
  \centering
  \includegraphics[width=0.50\textwidth]{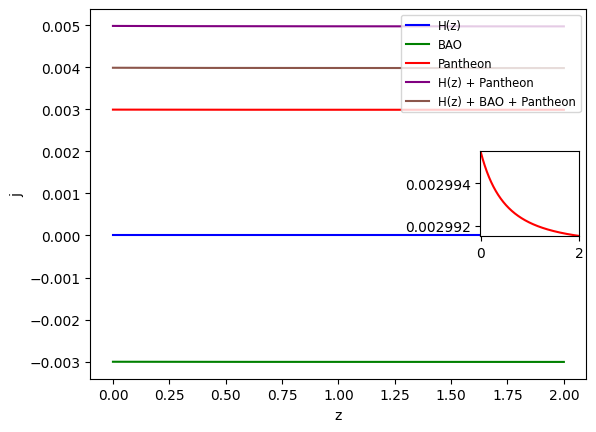}
  \caption{Visualisation of Jerk parameter $j$ vs $z$. Values of Jerk parameter at z = 0 are, for H(z) = 0.0 $s^{-3}$, for BAO = -0.0030045 $s^{-3}$, for Pantheon = 0.00299547 $s^{-3}$, for $H(z)_1 = H(z) + Pantheon = 0.00498738~s^{-3}$ and for $H(z)_2 = H(z) + Pantheon + BAO = 0.00399194~s^{-3}$}
  \label{fig:jerk}
\end{figure}
%%%%%%%%%%%%%%%%%%%%%%%%%%%%%%%%%%%%%%%%%%%%%%%%%%%%%%%%%%%%%%%%%%%%%%%%%%%%%%%%%%

\subsection{Jerk Parameter}
The Hubble parameter, the deceleration parameter, and the jerk parameter - all offer profound insights into cosmic evolution. The jerk parameter, denoted as $j$, measures how the universe's acceleration changes over time. ~\cite{Tripathy2021}. A positive $j$ ($j\ge0$) means that the universe's acceleration is speeding up, which means it is getting faster. On the other hand, a negative $j$ ($j\le0$) implies the acceleration is slowing down. Various dark energy theories predict different values for the jerk parameter. By measuring $j$ and comparing it to these predictions, scientists can get clues about the nature of the dark energy, which is supposed to be the mysterious force driving the universe's acceleration. When combined with other cosmic parameters, the jerk parameter helps improve our models of the universe. It rigorously tests these models, ensuring they accurately match what we observe in the real universe. For our model, we have obtained parameters, especially the variation of $j$ w.r.t the redshift $z$, which is shown in Fig. \ref{fig:jerk}.

\section{Conclusion} \label{sec6}
The purpose of the present investigation is basically to provide a mathematical model under the $f(R,T)$ modified gravity. The scheme to satisfy our motivation we have considered FRLW cosmology under a specific functional form $f(R,T) = R + \xi RT$ with the assumption of homogeneous and isotropic spacetime. We have constructed the Einstein field equation under the modified platform, i.e. $f(R,T)$ gravity, for homogeneous and isotropic spacetime and found the expressions for several cosmological parameters within the redshift range $0 \leq z \leq 1.965$. 

However, to obtain the model parameters $\alpha$, $\beta$ and $H_0$ we have uniquely employed the Markov Chain Monte Carlo (MCMC) method. The constrained values of the Hubble parameter in the present era are as follows: $H_0 = 67.098^{+2.148}_{-1.792}$ km s$^{-1}$ Mpc$^{-1}$, $H_0 = 67.588^{+2.229}_{-2.170}$ km s$^{-1}$ Mpc$^{-1}$, $H_0 = 66.270^{+2.215}_{-2.181}$ km s$^{-1}$ Mpc$^{-1}$, $H_0 = 65.960^{+2.380}_{-1.834}$ km s$^{-1}$ Mpc$^{-1}$, $H_0 = 66.274^{+2.015}_{-1.864}$ km s$^{-1}$ Mpc$^{-1}$. To verify the model, we have used the Hubble parameter ($H(z)$) dataset, Baryon Acoustic Oscillations (BAO) dataset, Pantheon dataset, combined $H(z)$ + Pantheon dataset and combined $H(z)$ + BAO + Pantheon dataset, respectively. For Pantheon dataset and analysis we have specifically followed the materials of the refs.~\cite{Popovic2021,Brout2022,Scolnic2022}.

In this connection, one point we would like to add here regarding the functional form of $f(R,T)$ Mishra et al.~\cite{Mishra2020a} argue that for any modified theory of gravity, we should modify Einstein's GR following the prescriptions, i.e. either via the geometrical part or the matter part with a positive energy density and negative pressure to demonstrate the cosmic acceleration~\cite{Mishra2020b}. Harko et al.~\cite{Harko2011} proposed the functional form of $f(R,T)$ as follows: (i) $f(R,T)=R+2f(T)$, (ii) $f(R,T)=f_1 (R)+f_2 (T)$ and (iii) $f(R,T)=f_1 (R)+f_2 (R)f_3 (T)$, where $f_1 (R)$, $f_2 (R)$, $f_2 (T)$, $f_3 (T)$ are arbitrary functions of their respective arguments. However, in the present investigation, we have specifically adopted the scheme $f(R,T)=f_1 (R)+f_2 (T)$.

It has been observed that the derived numerical value for $H_o$ corresponds to the results observed by the Plank collaboration group. We have analysed the model by studying the energy conditions as a physical test. In addition to this, we have also executed a few pathological examinations through the jerk parameter, the Om diagnostics and the state finder diagnostic tools. Our observations and findings, as exhibited via the Figs. 1--19 and Tables indicate that our model on the FRLW cosmological model is consistent as far as physically imposed boundary conditions. Altogether, within a specified range of constraints, the present model seems viable, exhibiting interesting features and attributes of cosmic spacetime.

% \newpage

\section*{Acknowledgements}
S. Ray acknowledges support from the Inter-University Centre for Astronomy and Astrophysics (IUCAA), Pune, India, under the Visiting Research Associateship Programme and facilities under ICARD, Pune at CCASS, GLA University, Mathura, India. L.K. Sharma is thankful to IUCAA for approving a short visit when the idea of the present work has been conceived.

\section*{Data Availability Statement}
In this manuscript, we have used observational data as available in the literature. [Authors' comment: Our work does not produce any form of new data.] 

\section*{Conflicts of Interest}
The authors assert that there are no conflicts of interest pertaining to the publication of this work. 

% \newpage
\section*{Appendix}
%%%%%%%%%%%%%%%%%%%%%%%%%%%%%%%%%%%%%%%%%%%%%%%%%%%%%%%%%%%%%%%%%%%%%%%%%%%%%%%%%%
\begin{table}[htbp]
  \caption{57 Point Hubble Data obtained from \cite{Sharav2017}}
  \label{tab:57datapt}
  \tiny
  \begin{tabular}{|>{\centering\arraybackslash}p{2cm}|>{\centering\arraybackslash}p{2cm}|>{\centering\arraybackslash}p{2cm}|}
    \hline
    \textbf{$H(z)$} & \textbf{$\sigma_H$} & \textbf{$z$} \\
    \hline
        69.0   & 19.6  & 0.07\\
        69.0   & 12.0  & 0.9\\
        68.6   & 26.2  & 0.12\\
        83.0   & 8.0   & 0.17\\
        75.0   & 4.0   & 0.1791\\
        75.0   & 5.0   & 0.1993\\
        72.9   & 29.6  & 0.2\\
        77.0   & 14.0  & 0.27\\
        88.8   & 36.6  & 0.28\\
        83.0   & 14.0  & 0.3519\\
        83.0   & 13.5  & 0.3802\\
        95.0   & 17.0  & 0.4\\
        77.0   & 10.2  & 0.4004\\
        87.1   & 11.2  & 0.4247\\
        92.8   & 12.9  & 0.4497\\
        89.0   & 34.0  & 0.47\\
        80.0   & 99.0  & 0.4783\\
        97.0   & 62.0  & 0.48\\
        104.0  & 13.0  & 0.593\\
        92.0   & 8.0   & 0.6797\\
        105.0  & 12.0  & 0.7812\\
        125.0  & 17.0  & 0.8754\\
        90.0   & 40.0  & 0.88\\
        117.0  & 23.0  & 0.9\\
        154.0  & 20.0  & 1.037\\
        168.0  & 17.0  & 1.3\\
        160.0  & 33.6  & 1.363\\
        177.0  & 18.0  & 1.43\\
        140.0  & 14.0  & 1.53\\
        202.0  & 40.0  & 1.75\\
        186.5  & 50.4  & 1.965\\
        79.69  & 2.99  & 0.24\\
        81.7   & 6.22  & 0.3\\
        78.18  & 4.74  & 0.31\\
        83.8   & 3.66  & 0.34\\
        82.7   & 9.1   & 0.35\\
        79.94  & 3.38  & 0.36\\
        81.5   & 1.9   & 0.38\\
        82.04  & 2.03  & 0.4\\
        86.45  & 3.97  & 0.43\\
        82.6   & 7.8   & 0.44\\
        84.81  & 1.83  & 0.44\\
        87.9   & 2.03  & 0.48\\
        90.4   & 1.9   & 0.51\\
        94.35  & 2.64  & 0.52\\
        93.34  & 2.3   & 0.56\\
        87.6   & 7.8   & 0.57\\
        96.8   & 3.4   & 0.57\\
        98.48  & 3.18  & 0.59\\
        87.9   & 6.1   & 0.6\\
        97.3   & 2.1   & 0.61\\
        98.82  & 2.98  & 0.64\\
        97.3   & 7.0   & 0.73\\
        224.0  & 8.6   & 2.3\\
        224.0  & 8.0   & 2.33\\
        222.0  & 8.5   & 2.34\\
        226.0  & 9.3   & 2.36\\ 
    \hline
  \end{tabular}
\end{table}
\newpage
%%%%%%%%%%%%%%%%%%%%%%%%%%%%%%%%%%%%%%%%%%%%%%%%%%%%%%%%%%%%%%%%%%%%%%%%%%%%%%%%%%


\newpage
\begin{thebibliography}{50}
% \section*{References}

\bibitem{Hubble1929} E. Hubble, A Relation between Distance and Radial Velocity among Extra-Galactic Nebulae, Proc. Nat. Acad. Sci., {\bf 15}, 168--173 (1929)

\bibitem{Riess2004} A.~C.~Becker et al., Exploring the Outer Solar System with the ESSENCE Supernova Survey, Astrophys. J. Lett. \textbf{682}, L53 (2008).???

\bibitem{Riess1998} A. G. Riess et al.  Astron. J. \textbf{116}, 1009 (1998).

\bibitem{Perlmutter1998} S.~Perlmutter et al. [Supernova Cosmology ProUniverseiscovery of a supernova explosion at half the age of the Universe and its cosmological implications, Nature \textbf{391}, 51-54 (1998).

\bibitem{Perlmutter1999} Supernova Cosmology Project collaboration, S. Perlmutter et al., Astrophys. J., \textbf{517}, 565 (1999).

\bibitem{Bennett2003} C. L. Bennett et al., Astrophys. J. Suppl. Ser. \textbf{148}, 1 (2003).	

\bibitem{Spergel2003} D. N. Spergel et al. Astrophys. J. \textbf{148}, 175 (2003).

\bibitem{Spergel2007} D. N. Spergel et al. Astrophys. J. \textbf{170}, 377 (2007).

\bibitem{Percival2010} W. J. Percival et al., Mon. Not. R. Astron. Soc. \textbf{401}, 2148 (2010).

\bibitem{Ade2014} P. A. R. Ade et al. (BICEP2 Collaboration), Phys. Rev. Lett. \textbf{112}, 241101 (2014).

\bibitem{Aghanim2018}A.~Gorce, M.~Douspis, N.~Aghanim and M.~Langer, Observational constraints on key-parameters of cosmic reionisation history, Astron. Astrophys. \textbf{616}, A113 (2018).

\bibitem{Mishra2019}B.~Mishra, P.~P.~Ray and R.~Myrzakulov, Bulk viscous embedded hybrid dark energy models, Eur. Phys. J. C \textbf{79}, 34 (2019).

\bibitem{Peebles2003} P.~J.~E.~Peebles and B.~Ratra, The Cosmological Constant and Dark Energy, Rev. Mod. Phys. \textbf{75}, 559-606 (2003).

\bibitem{Copeland2006} E.~J.~Copeland, M.~Sami and S.~Tsujikawa, Dynamics of dark energy, Int. J. Mod. Phys. D \textbf{15}, 1753-1936 (2006).

\bibitem{Caldwell2002} R.~R.~Caldwell, A Phantom menace?, Phys. Lett. B \textbf{545}, 23-29 (2002).

\bibitem{Kumar:2010vpj} S.~Kumar and C.~P.~Singh, Anisotropic dark energy models with constant deceleration parameter, Gen. Rel. Grav. \textbf{43}, 1427-1442 (2011).

\bibitem{Yadav:2011bj} A.~K.~Yadav and B.~Saha, LRS Bianchi-I Anisotropic Cosmological Model with Dominance of Dark Energy, Astrophys. Space Sci. \textbf{337}, 759-765 (2012).

\bibitem{Mishra:2017gtk} B.~Mishra, S.~K.~Tripathy and P.~P.~Ray, Bianchi-V string cosmological model with dark energy anisotropy, Astrophys. Space Sci. \textbf{363}, 86 (2018).

\bibitem{Mishra:2015jja} B.~Mishra and S.~K.~Tripathy, Anisotropic dark energy model with a hybrid scale factor, Mod. Phys. Lett. A \textbf{30}, no.36, 1550175 (2015).

\bibitem{Ali:2015mov} A.~T.~Ali, A.~Kumar Yadav and A.~K.~Alzahrani, Similarity dark energy models in Bianchi type - I space-time, Eur. Phys. J. Plus \textbf{131}, 415 (2016).

\bibitem{Goswami:2016ppx} G.~K.~Goswami, A.~K.~Yadav and R.~N.~Dewangan, Magnetised Strings in \ensuremath{\Lambda}-Dominated Anisotropic Universe, Int. J. Theor. Phys. \textbf{55}, 4651-4664 (2016).

\bibitem{Yadav:2016zvc} A.~K.~Yadav, A transitioning universe with anisotropic dark energy, Astrophys. Space Sci. \textbf{361}, no.8, 276 (2016).

\bibitem{Zaeem-ul-HaqBhatti:2016moh} M.~Z. Bhatti, Shear-free stellar solutions in $\Lambda$-dominated era, Eur. Phys. J. Plus \textbf{131}, 428 (2016).

\bibitem{Yousaf:2017} Z.~Yousaf, Stellar filaments with Minkowskian core in the Einstein- $\Lambda$ gravity, Eur. Phys. J. Plus \textbf{132}, no.6, 276 (2017).

\bibitem{Harko2011} T.~Harko, F.~S.~N.~Lobo, S.~Nojiri and S.~D.~Odintsov, $f(R,T)$ gravity, Phys. Rev. D \textbf{84}, 024020 (2011).

\bibitem{Zubair:2016} M.~Zubair, S.~Waheed and Y.~Ahmad, Static spherically symmetric wormholes in f(R, T) gravity, Eur. Phys. J. C \textbf{76}, no.8, 444 (2016).

\bibitem{Moraes:2016} P.~H.~R.~S.~Moraes, R.~A.~C.~Correa and R.~V.~Lobato, Analytical general solutions for static wormholes in $f(R,T)$ gravity, JCAP \textbf{07}, 029 (2017).

\bibitem{Das:2016} A.~Das, F.~Rahaman, B.~K.~Guha and S.~Ray, Compact stars in $f(R,\mathcal {T})$ gravity, Eur. Phys. J. C \textbf{76}, 654 (2016).

\bibitem{Singh:2015} V.~Singh and C.~P.~Singh, Friedmann Cosmology with Matter Creation in Modified $f(R,T)$ Gravity, Int. J. Theor. Phys. \textbf{55}, 1257-1273 (2016).

\bibitem{Myrzakulov:2012} R.~Myrzakulov, FRW Cosmology in $F(R,T)$ gravity, Eur. Phys. J. C \textbf{72}, 2203 (2012).

\bibitem{Houndjo:2011} M.~J.~S.~Houndjo, Reconstruction of $f(R,T)$ gravity describing matter dominated and accelerated phases, Int. J. Mod. Phys. D \textbf{21}, 1250003 (2012).

\bibitem{Das2016} A. Das, F. Rahaman, B.K. Guha and S. Ray, Compact stars in $f(R,T)$ gravity, Eur. Phys. J. C \textbf{76}, 654 (2016). 

\bibitem{Das2017} A. Das, S. Ghosh, B.K. Guha, S. Das, F. Rahaman and S. Ray, Gravastars in $f(R,T)$ gravity, Phys. Rev. D \textbf{95}, 124011 (2017). 

\bibitem{Deb2018a} D. Deb, F. Rahaman, S. Ray and B.K. Guha, Anisotropic strange stars under simplest minimal matter-geometry coupling in the $f(R,T)$ gravity, Phys. Rev. D \textbf{97}, 084026 (2018). 

\bibitem{Deb2018b} D. Deb, S. V. Ketov, S. K. Maurya, M. Khlopov, P. H. R. S. Moraes and S. Ray, Exploring physical features of anisotropic strange stars beyond standard maximum mass limit in $f(R,T)$ gravity,  Mon. Not. R. Astron. Soc. \textbf{485}, 5652 (2018). 

\bibitem{Biswas2019} S. Biswas, S. Ghosh, B. K.Guha, F. Rahaman and S. Ray, Strange stars in Krori-Barua space-time under $f(R,T)$ gravity,  Ann. Phys. \textbf{401}, 1 (2019). 

\bibitem{Deb2019a} D. Deb, S. V. Ketov, M. Khlopov and S. Ray, Study on charged strange stars in $f(R,T)$ gravity, J. Cosmol. Astropart. Phys. \textbf{10}, 070 (2019). 

\bibitem{Deb2019b} D. Deb, F. Rahaman, S. Ray and B.K. Guha, Strange stars in $f(R,T)$ gravity, J. Cosmol. Astropart. Phys. \textbf{03}, 044 (2019). 

\bibitem{Biswas2020} S. Biswas, D. Shee, S. Ray and B.K. Guha, Anisotropic strange star with Tolman-Kuchowicz metric under $f(R,T)$ gravity, Eur. Phys. J C \textbf{80}, 175 (2020).

\bibitem{Hulke2020} N. Hulke, G. P. Singh, B. K. Bishi, A. Singh, Variable Chaplygin gas cosmologies in $f(R,T)$ gravity with particle creation, New Astron. \textbf{77}, 101357 (2020).

\bibitem{Maurya2020a} D. C. Maurya, Modified $F(R,T)$ cosmology with observational constraints in Lyra's geometry, Int. J. Geom. Meth. Mod. Phys. \textbf{17}, 2050001 (2020).
		
\bibitem{Maurya2020b} D. C. Maurya, A. Pradhan, A. Dixit, Domain walls and quark matter in Bianchi type-V universe with observational constraints in $F(R,T)$ gravity, Int. J. Geom. Meth. Mod. Phys. \textbf{17}, 2050014 (2020). 
		
\bibitem{Maurya2020c} D. C. Maurya, Transit cosmological model with specific Hubble parameter in $F(R,T)$ gravity, New Astron. \textbf{77}, 101355 (2020). 

\bibitem{Biswas2021} S. Biswas, D. Deb, S. Ray, B.K. Guha, ``Anisotropic charged strange stars in Krori-Barua spacetime under $ f(R,T)$ gravity, Ann. Phys. \textbf{428}   168429 (2021). 

\bibitem{Maurya2021} S.K. Maurya, F. Tello-Ortiz and S. Ray, ``Decoupling gravitational sources in $f(R,T)$ gravity under class I spacetime, Phys. Dark Univ. \textbf{31}  100753 (2021). 

\bibitem{Maurya2023} D. C. Maurya, J. singh, L. K. Gaur, Dark Energy Nature in Logarithmic $f(R,T)$ Cosmology, Int. J. Geom. Meth. Mod. Phys. \textbf{20}, 2350192 (2023). 

\bibitem{Moraes2017} P.~H.~R.~S.~Moraes and P.~K.~Sahoo, The simplest non-minimal matter-geometry coupling in the $f(R,T)$ cosmology, Eur. Phys. J. C \textbf{77}, 480 (2017).

\bibitem{Sharma:2019hqe} L.~K.~Sharma, B.~K.~Singh and A.~K.~Yadav, Viability of Bianchi type V universe in $f(R,T)= f_{1}(R)+f_{2}(R)f_{3}(T)$  gravity, Int. J. Geom. Meth. Mod. Phys. \textbf{17}, 2050111 (2020).

\bibitem{Sharma:2018ikm} L.~K.~Sharma, A.~K.~Yadav, P.~K.~Sahoo and B.~K.~Singh, Non-minimal matter-geometry coupling in Bianchi I space-time, Res. Phys. \textbf{10}, 738-742 (2018).

\bibitem{Yadav:2010ge} A.~K.~Yadav, Thermodynamical Behaviour of Inhomogeneous Universe with Varying Lambda in Presence of Electromagnetic Field, Int. J. Theor. Phys. \textbf{49}, 1140-1154 (2010).

\bibitem{Bhardwaj2019} V. K. Bhardwaj, M. K. Rana, A. K. Yadav, Bulk viscous Bianchi-V cosmological model within the formalism of $f(R,T)=f_1 (R)+f_2 (R)f_3 (T)$ gravity, Astrophys. Space Sc. {\bf 364}, 136 (2019).

\bibitem{Kumar2012} S.~Kumar, Observational constraints on Hubble constant and deceleration parameter in power-law cosmology, Mon. Not. R. Astron. Soc. \textbf{422}, 2532-2538 (2012).

\bibitem{Kumar:2010si} S.~Kumar, Some FRW Models of Accelerating Universe with Dark Energy, Astrophys. Space Sci. \textbf{332}, 449-454 (2011).

\bibitem{Yadav:2010dj} A.~K.~Yadav, F.~Rahaman and S. Ray, Dark energy model with variable equation of state parameter, Int. J. Theor. Phys. \textbf{50}, 871-881 (2011).

\bibitem{Yadav:2011dr} A.~K.~Yadav, Some Anisotropic Dark Energy Models in Bianchi Type-V Space-time, Astrophys. Space Sci. \textbf{335}, 565-575 (2011).

\bibitem{Yadav:2010ah} A.~K.~Yadav and L.~Yadav, Bianchi Type III Anisotropic Dark Energy Model with Constant Deceleration Parameter, Int. J. Theor. Phys. \textbf{50}, 218-227 (2011).

\bibitem{dec1} A. Feinstein and J. lbanez, Class. Quantum Grav. \textbf{10}, L227 (1993). 

\bibitem{rayc1} A. K. Raychaudhuri, Theoretical Cosmology, Oxford University Press (First Edition, 1979). 

\bibitem{Akarsu2019} {\"O}. Akarsu, S. Kumar, S. Sharma and L. Tedesco, Constraints on a Bianchi type I spacetime extension of the standard $\Lambda$ CDM model, Phys. Rev. D \textbf{100}, 023532 (2019).

\bibitem{Singh2015} T. Singh, R. Chaubey, A. Singh, Bounce conditions for FRW models in modified gravity theories, Eur. Phys. J. Plus \textbf{130}, 31 (2015).

\bibitem{Singh2023} A. Singh, Homogeneous and anisotropic cosmologies with affine EoS: a dynamical system perspective, Eur. Phys. J. C \textbf{83}, 696 (2023).

\bibitem{Sahni2008} V. Sahni, A. Shafieloo and A. A. Starobinsky, Two new diagnostics of dark energy, Phys. Rev. D {\bf 78} 103502 (2008).

\bibitem{Tripathy2021} S. K. Tripathy, B. Mishra, M. Khlopov and S. Ray, Cosmological models with a hybrid scale factor, Int. J. Mod. Phys. D  {\bf 30} 2140005 (2021).

\bibitem{Mandal2022} S. Mandal, A. Singh, R. Chaubey, Observational constraints and cosmological implications of NLE model with variable $G$, Eur. Phys. J. Plus \textbf{137}, 1246 (2022). 

\bibitem{Myrzakulov2023} N. Myrzakulov, M. Koussour and A. Mussatayeva, Quintessence-like features in the late-time cosmological evolution of f(Q) symmetric teleparallel gravity, Chin. J. Phys. {\bf 85}, 345--358 (2023).

\bibitem{Shahalam2015} M. Shahalam, S. Sami, A. Agarwal, Om diagnostic applied to scalar field models and slowing down of cosmic acceleration, Mon. Not. R. Astron. Soc. {\bf 448}, 2948 (2015).

\bibitem{Mishra2020a} B. Mishra, S.K. Tripathy and S. Ray, Cosmological models with squared trace in modified gravity, Int. J. Mod. Phys. D \textbf{29},  2050100 (2020). 

\bibitem{Mishra2020b} B. Mishra, F. MD. Esmaeili, S. Ray, Cosmological models with variable anisotropic parameter in $f(R,T)$ gravity, Ind. J. Phys. {\bf 95}, 2245 (2021).

\bibitem{Sharav2017} G. S. Sharov, S. Bhattacharya, S. Pan, R. C. Nunes and S. Chakraborty, A new interacting two-fluid model and its consequences, Mon. Not. R. Astron. Soc. \textbf{466}, 3497-3506 (2017).

\bibitem{Popovic2021} B. Popovic, D. Brout, R. Kessler and D. Scolnic, The Pantheon+ Analysis: Forward-Modeling the Dust and Intrinsic Colour Distributions of Type Ia Supernovae, and Quantifying their Impact on Cosmological Inferences, arXiv:2112.04456 [astro-ph.CO].

\bibitem{Brout2022} D. Brout et al., The Pantheon+ Analysis: SuperCal-fragilistic Cross Calibration, Retrained SALT2 Light-curve Model, and Calibration Systematic Uncertainty, Astrophys. J. {\bf 938}, 111 (2022).

\bibitem{Scolnic2022} D. Scolnic et al., The Pantheon+ Analysis: The Full Data Set and Light-curve Release, Astrophys. J. {\bf 938}, 113 (2022).


\end{thebibliography}
\end{document}